# DISCOVERY OF DAEMONS MAKES
# POWER GENERATION BY DAEMON-ASSISTED CATALYSIS OF LIGHT NUCLEI FUSION IN A BALL LIGHTNING A REALITY

E. M. Drobyshevsky

*Ioffe Physico-Technical Institute, Russian Academy of Sciences, 194021 St.Petersburg, Russia*
*emdrob@mail.ioffe.ru*

***Abstract.*** In 1997, we proposed a model of the ball lightning (BL) whose activity is accounted for by energy release in the fusion of light nuclei, most probably, carbon in organic fibers (Proc. ISBL 97, p.157). The fusion is provided by catalytic action of superheavy negative particles making up the Galactic Dark Matter. We called them DArk Electric Matter Objects, or daemons. The daemons are assumed to be elementary black holes ($M \sim 10^{-5}$ g) carrying a charge of up to $Ze = 10e$. Experiments have culminated in 2000 by the discovery of daemons. We used the two-screen scintillation technique with a scintillator ZnS(Ag). Measurements showed the daemon flux striking the Earth to be $\sim 10^{-9}$ cm$^{-2}$s$^{-1}$ for an object velocity of as low as $\sim<10-30$ km/s. The half-year periodicity of the flux was revealed, which can be assigned to daemons being captured into helio- and geocentric orbits as the Solar system moves through the DM background (see *astro-ph*/0402367 in http://www.arXiv.org).

The next step in creating a daemon-mediated BL to achieve *controlled energy generation* through catalytic fusion of carbon nuclei will be a search for methods of capture and stable holding of a daemon, which could involve the use of the tip of a carbon fiber, as it apparently occurs in nature.

### I. Introduction. General considerations on the nature of the BL

It is believed that the phenomenon called the BL can on different occasions have different, chemical or plasma, nature. All these manifestations may be described by an eyewitness as the BL. No objections can be raised against this. But neither of these approaches can explain the energy of $\sim 1\text{-}10^3$ MJ released (in an explosive way) in some of the well-documented events. This energy is comparable to that of a lightning (up to $\sim 10^3\text{-}10^4$ MJ) while exceeding by far the chemical or plasma energy that can be contained in the visible volume of a BL. On these grounds, Barry [1], for instance, believes such reports and the estimates based on them to be simply improbable. The only consideration that could justify this approach is an *a priori* belief that the BL phenomenon which had defied explanation for centuries still can be interpreted based on the already known relations and laws. While this approach appears reasonable enough, one can readily perceive in it the vicious circle logic.

In 1996-1997, we proposed a novel hypothesis of the BL phenomenon, which is capable of accounting for virtually all of its known properties, conceivably even $\sim 10^3$ MJ explosions. It is based on modern cosmological concepts postulating that the Universe started from Planckian scales, as well as on the well established fact that the major part of the mass contained in the galaxies (up to 90%) consists of the so-called Dark Matter. Despite considerable efforts of researchers, the nature of the DM has been remaining a mystery until recently. Following a number of authors [2-4], we assumed [5] that the DM is made up of Planckian elementary black holes ($M \approx 2\text{-}3\times 10^{-5}$ g, $r_g \approx 3\text{-}2\times 10^{-33}$ cm) carrying also an electric charge ($Ze \approx 10e$), whose repulsive action is counterbalanced by the gravitational field. The concentration of such DArk Electric Matter Objects, daemons, in the Galactic halo is $\sim 10^{-19}$ cm$^{-3}$ for a velocity dispersion of $\sim 200\text{-}300$ km/s. There should also be a Galactic disk population rotating about the center of the Galaxy. The velocity dispersion of DM objects in it was estimated by Bahcall *et al* [6] to be 4-30 km/s. In passing through the Sun, some daemons slow down somewhat and are captured first into elongated orbits with a perihelion within the Sun. The resistance exerted by the matter makes the orbits contract, with the daemons accumulating inside the Sun. If a daemon moving in an elongated orbit crosses the Earth's orbit and traverses its gravitational sphere, the Earth transforms eventually the daemon's orbit to quasi-circular.

Heavy negative particles are capable of catalyzing fusion of the nuclei of light elements. For an illustration, recall the muon-mediated catalysis of deuteron fusion, the process which had been believed to offer an alternative to future bulky hot-plasma machines for producing energy. The muons live, however, only 2.2 μs and can catalyze in this time not more than ~100 fusion events.

Considered from this standpoint, negative daemons have indisputable assets. They are massive, stable, and, having a large charge, are capable of catalyzing fusion of nuclei up to oxygen. The catalysis of fusion of carbon nuclei by a daemon that entered a thin carbon fiber is the key element of our BL hypothesis [5, 7]. The attendant phenomena can account for the BL properties. We are going to recall this hypothesis in Sec. II.

In 1996-2004, we undertook considerable efforts to detect daemons by various methods. In 2000, first promising results were obtained [8]. This work has been continuing successfully but regrettably slowly since then (detected events build up not so fast as we would like because of the astronomical nature of the objects), so that our understanding of the properties of these particles and of the implications of their existence is progressing. The main results obtained in the detection of daemons will be covered in Sec. III.

The discovery of daemons provides a reliable physical foundation for our BL hypothesis. The major implication of the existence of a daemon BL is the prospect of developing a clean and compact source of nuclear power with an output of 1-10 kW or more. We have been discussing this subject as far back as 1996-1997. There are now solid grounds for conducting research in this promising area (see Conclusion).

**II. Daemon energy release modes and the nature of the BL**

Our ideas concerning the interaction of daemons with matter have undergone since 1996-1997 a considerable progress. We came to the understanding that the principal mode of daemon-mediated energy release in nature is not the catalysis of fusion of light nuclei, a process we thought in 1996 is realized in the Sun [9], but rather that of proton decay. Indeed, capture by a daemon, say, of a Fe nucleus releases an energy $W = 1.8ZZ_nA^{-1/3}$ MeV ≈ 120 MeV, which results in emission of ~12 nucleons and their clusters. The ground state radius of the daemon in the remainder of the nucleus (with $Z_n > 24/Z$) is less than that of the proton, so that the daemon captured by a nucleus resides actually inside the proton. Nature did not create forces stronger than the Planckian ones. Therefore, while the details of this process cannot be established presently, it appears plausible that the daemon catalyzes proton decay [10, 11]. One could suggest here an analogy with the monopole-assisted decay of the proton, which was proposed by Rubakov (he estimated the decay time as $10^{-6\pm2}$ s) [12]. As follows from our experiments (see Sec. III below), the daemon-stimulated proton decay time $\Delta\tau_{ex} \approx 10^{-6}$ s. Such a process can account for one half of the heat flux (≈20 TW) and the $^3$He flux out of the Earth, part of the energetics, and emission of non-electron neutrinos from the Sun, positron generation in the Galactic nucleus [13], and other observations. At the same time, the power liberated by the daemon in proton disintegration does not exceed $m_pc^2/\Delta\tau_{ex}$ = 0.15 mW, a figure inconsistent with usual ideas of the BL.

Daemon-mediated catalysis of the fusion of light nuclei, particularly of carbon, could be more efficient. This is associated with the large cross section of the process at MeV energies, so that fusion of daemon-captured nuclei should occur in $\sim10^{-20\pm2}$ s [14]. Therefore, energy release can depend on the capture rate of nuclei rather than on the rate of their fusion. In graphite, the capture rate of $C^{4+}$ ions is as high as $\sim2\times10^{15}$ s$^{-1}$ [5], and the energy release, $\sim10^2$-$10^3$ W, a figure well consistent with a stationary BL.

One could conceive of the following scenario of the formation and evolution of a BL. On entering a carbon fiber with a diameter ~1-10 μm, the daemon starts to capture carbon nuclei and catalyze their fusion, $^{12}C + ^{12}C \rightarrow ^{24}Mg$ + 13.9 MeV. Capture events with energies measured in many MeV eject Auger and internal conversion electrons (with energies of up to $10^4$-$10^5$ keV). In passing through the fiber material, they drag along up to $\sim10^2$-$10^3$ new electrons. As a result, the fiber charges positively, and the electrons surround it with a transparent luminous halo of excited air ~10 cm in radius. The deexcited $^{24}$Mg compound nuclei ejected out of the daemon neighborhood by internal conversion slow down in a few mm of the air to produce an internal core well visible in a BL. The positive charge of the fiber generates an electric field of ~100 V/nm at its tip where the daemon glides to, which is high enough to confine the daemon to the fiber. The heavy daemon at the fiber tip orients the fiber with its tip down. The charge streaming down from the tip creates an "electric wind" making it possible for the fiber to levitate and, if it is asymmetric, to float horizontally, sometimes even against a weak wind, although usually a light fiber would be dragged by air motion. The BL phenomenon lasts as long as new carbon nuclei are available for the daemon to capture, and the conditions are favorable for the latter to keep suspended electrically in the fiber. For an energy release of ~$10^2$ W, about ~$2 \times 10^{-8}$ cm$^{-3}$ carbon is converted to magnesium per second, which for a 10-μm-thick fiber corresponds to its length of ~0.2 mm. These figures sound quite reasonable for particles of soot. In the case of long organic fibers, the BL lifetime may be measured in many seconds. The cases of rolling along horizontal surfaces, sometimes interrupted by hopping to leave a charred trace can be readily accounted for by liberation of gas from the surface under heating. Such processes force the daemon-containing dust particle to rise, deflect, end so on.

The daemon model is capable of accounting for all the known properties of the quiet BL. In explaining its behavior, we do not step beyond the concepts of conventional nuclear physics and the processes of catalysis of light nucleus fusion known from muon physics.

As for the explosive BL manifestations, the highest energies released here are also comparable to the daemon energetics, namely, $Mc^2 \approx 2000$ MJ. It is conceivable that a daemon coming in contact with coherent structures of matter (like magnetic domains in iron) with a mass ~$M$ may reveal its quantum relativistic properties. One should not exclude the possibility of its absorbing these structures, replicating etc. At any rate, in all cases of explosion of a BL one may suspect its encounter with iron.

It thus follows that the BL is a progenitor of a compact generator of nuclear energy which employs the potential of the new physics prompted by cosmology.

### III. Experimental detection of daemons

Our experiments aimed at detection of negative daemons offer a solid footing for the above BL hypothesis (although the initial impetus to our study of daemons was provided by an analysis of the BL properties; the very first our experiments dealing with the nature of the BL assumed the possibility of phase transitions in superdense plasma [15]).

The assumptions underlying our search for daemons were not typical of the standard experiments on detection of DM objects:

1. Operating on the premise that the daemon flux from the center of the Galaxy moving with a velocity of ~250 km/s is ~$10^{-12}$ cm$^{-2}$s$^{-1}$, we, unlike the other groups of scientists, focused attention on a search for slow (~10-30 km/s) objects that had been accumulating in heliocentric orbits. By our optimistic estimates, their flux can reach ~$3 \times 10^{-7}$ cm$^{-2}$s$^{-1}$ [5].

2. Daemons are nuclear active particles:

(a) First, they are capable of catalyzing fusion of light nuclei. This is why we made attempts at constructing detectors of Li or Be plates. These experiments did not, however, yield a positive result [16];

(b) Second, the binding energy of a daemon to, say, a Zn nucleus is $W \approx 130$ MeV, which in the case of the capture of a nucleus would result in ejection of atomic electrons and of up to ~12 nucleons capable of initiating a scintillation. This accounts for the advantages of ZnS(Ag) (light yield up to 28% of absorbed energy) compared to the organic scintillators (light yield ≤3-4%; besides, for the carbon nucleus $W \approx 47$ MeV, while the excitation energy of the first level in $^{12}$C is 4.4 MeV, which is larger than ~1 MeV for Zn, so that capture of $^{12}$C is less probable than that of Zn).

(c) Third, a daemon, if finding itself in a proton, decomposes it in a short time. As a result, with the passage of time the daemon "digests" the captured nucleus and becomes capable of capturing a new nucleus.

As follows from items 1 and 2, in contrast to fast cosmic ray particles whose mean free path between interactions with nuclei in condensed matter ~10 cm, for the slow (~10 km/s) negatively charged daemons this length reduces to ~$10^{-3}$ cm. The reason for this is that during the fairly long time the daemon needs to fly past a nucleus the latter will shift towards the daemon. Therefore, it appears preferable to use for daemon detection thin (~10 µm thick) scintillators with a large enough atomic weight of the type of ZnS(Ag) and the like. They are suitable for detection of daemons (to be precise, of protons from the nuclei captured by them) while being only weakly sensitive to the background and conventional cosmic radiation (indeed, MeV-range α-particles and protons leave a sizable part of their energy in such a layer whereas the β and γ radiation will loose less than 10 keV). Therefore, in the first stage of experimentation one should not even worry about protecting the detector against the radiation background.

These principles served as a basis in designing a detector made up of four modules [8]. Each module consisted of two 0.5×0.5 $m^2$ transparent polystyrene plates 4 mm thick, which were coated on the underside by 3.5 mg/$cm^2$ layer of ZnS(Ag) powder. Both screens were arranged horizontally at a distance of 7 cm from one another at the center of a tinned-iron cubic case (0.3-mm-thick Fe sheet coated on both sides with 2 µm Sn layer), 51 cm on a side. The top lid of the case was of black paper. Black paper separated also the screens. Each plate was viewed with its PM tube from a distance of 22 cm. The PM output signals were fed to a double-trace digital oscilloscope and fixed during ±100 µs from the trigger pulse. The trigger signal was taken from the top PM tube. If the second trace also had a signal, the event was recorded in computer memory. Double events with zero time shifts (frequently recorded in more than one module simultaneously) were assigned to cosmic rays.

The signals observed were of two types. The first type, with a short leading edge (≤1 µs) and a trailing edge determined by the PM tube load, is typical of cosmic rays and intrinsic multi-electron PM tube noise. We called them Noise-Like Scintillations (NLSs).

The second type, Heavy Particle Scintillations (HPSs), is characteristic of scintillations produced by α-particles. Their oscilloscope traces have a smooth maximum at 2.5-3 µs from the beginning of the signal, so that their integrated area normalized against the amplitude is ~ twice that of the NLSs.

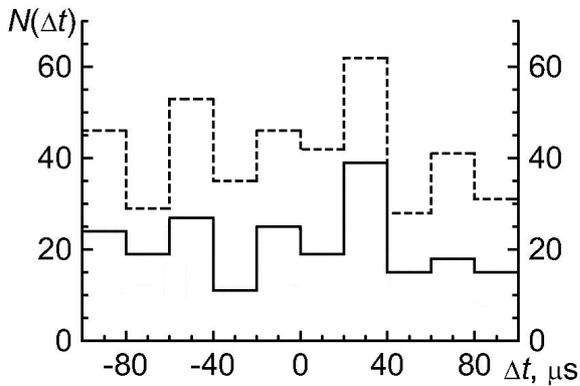

Fig. 1. (---) Distribution $N(\Delta t)$ of pair scintillation events on their time shift (relative to the upper channel events). (—) Similar distribution which takes into account only HPS type events at the upper channel.

Our purpose was to find statistically correlated signals with a time shift corresponding to particles with velocities ≤30 km/s. And we found such signals. Fig. 1 plots a distribution $N(\Delta t)$ of paired signals in their time shift for March, 2000 ($\Delta t > 0$ corresponds to the bottom screen signal coming later, i.e., to the object moving downward) [8,13].

The daemon concept allows an uncontradictory interpretation of the $N(\Delta t)$ pattern. One clearly sees a maximum at $20 < \Delta t < 40$ µs corresponding to a flux of ~$10^{-9}$ $cm^{-2}s^{-1}$. Using only the events with an HPS on the first trace, its confidence level increases to 99.5%, despite the decrease in the total number of events from 413 to 212. This suggests that the maximum is not a consequence of interference of some kind, and that the scintillations are initiated by protons ejected in the daemon-mediated capture of a Zn (or S) nucleus. The shift by $\Delta t \approx 30$ µs can be accounted for if we assume that the daemon, in its downward motion, crosses the bottom scintillator without interacting with it, because it is still "poisoned" by the Zn (or S) nucleus remainder. In propagating further down while continuing its digestion, one proton after another, the daemon reduces the charge of the nuclear remainder down to $Z_n \leq 9$ by the time it reaches the bottom lid of the case ~20-30 µs thereafter. At this instant, the daemon captures a new Sn or Fe nucleus with emission of protons and electrons (Auger, internal conversion, and refilling). Reaching the bottom scintillator, protons with $E \geq 3.9$ MeV and electrons excite in it a time-shifted scintillation.

We can draw two important conclusions:

(1) The peak with $\Delta t \approx 20\text{-}40$ μs can be identified with the velocity ~10-15 km/s, which corresponds to objects falling from the Near-Earth Almost-Circular Heliocentric Orbits (NEACHOs), where the daemons are transferred by combined gravitational action of the Sun and the Earth. These orbits cross the Earth's orbit in the regions of the shadow and antishadow of the Sun, which it creates in moving relative to the daemon population of the Galactic disk (see also below);

(2) A comparison of the numbers of protons in the Zn nucleus evaporated and disintegrated in its interaction with the daemon provides an estimate of the proton decay time of ~$10^{-6}$ s.

An analysis of fine features in the $N(\Delta t)$ distribution is far from a simple problem. The situation is complicated also by the fact that after a year of various check experiments including variation of the detector parameters, we understood that, in addition to a noticeable effect seemingly small changes in detector design exert on the pattern of $N(\Delta t)$, there is a strong seasonal variation in the daemon flux. A three-year exposure of the four-module detector maintained unchanged (although it was clear that its parameters needed improving) showed convincingly that the deviation of $N(\Delta t)$ from *const* has a period of 0.5 yr and passes through a maximum in January-March and July-September [13], thus defining the direction which roughly coincides with the Sun's apex relative to the stars.

An analysis of the $N(\Delta t)$ distribution shows the existence of an upward daemon flux. In particular, the well-pronounced minimum at $\Delta t \approx -30$ μs can be explained as due to the daemon's not having had digested in 30 μs a Sn nucleus captured in crossing the bottom lid in the case.

An essential observation is the manifestation of a daemon population with a velocity of only ~5 km/s ($|\Delta t| > 60$ μs). This means that part of the NEACHO's daemons is captured into Geocentric Earth-Surface-Crossing Orbits (GESCOs). As a result of the resistance the GESCO daemons meet in crossing the Earth's material, they slow down gradually to build up eventually in a central kernel to a mass of up to $10^{19}$ g and a few cm in size [17]. In their last emergencies above the Earth's surface, the velocity of the daemons may drop down to a few m/s (cm/s?) or less, a feature essential for understanding properly the BL phenomenon.

## IV. Prospects of realizing a BL and a compact nuclear power generator

It is hard to grasp the staggering implications of the discovery of the daemon, most likely, a Planckian particle. We are entering the realm of energies ~$10^{19}$ GeV ≈ 2000 MJ, a realm unimaginable thus far for the physics of elementary particles. This is the scale from which the Universe and the matter itself began their existence.

Practical consequences of the discovery of daemons are more tangible. Catalysis of the fusion of light nuclei offers a possibility of building a compact source of ecologically clean nuclear power. Actually, Nature demonstrates it to the Man in the form of a BL. What major stages in the realization of this idea could be visualized?

It appears that one will have to address the following two major issues:

First, one has to slow down the daemon. Because of the giant mass of the daemon, this is a problem of daunting complexity. Indeed, for $Ze = 10e$, a potential difference of 10 MV is capable of stopping the daemon moving with a velocity of only 3-4 cm/s. One would have to search initially for daemons which have already been decelerated by the Earth in their last emergencies above its surface. It is quite possible that their kinematics is affected noticeably in thunderstorm conditions by pulses of atmospheric electricity. Thus, detection of superslow daemons with $V \leq 1\text{-}0.01$ km/s would be a promising approach. Using the phenomenon of daemon-stimulated proton decay seems to make it a not very difficult task. Traversal by a slow daemon of a massive enough block of a CsI(Tl)-type scintillator will be signaled by the appearance of multitooth scintillations of similar amplitudes corresponding to ~$m_p c^2 = 938$ MeV. These teeth are produced by consecutive proton decays occurring with an average interval of ~1 μs in the remainder of the captured heavy nucleus (Cs or I). In its turn, the neutron-excessive remainder must also emit a great deal of a somewhat delayed neutrons.

Detection of such events would be already a landmark in science. On the other hand, this would be a no-loss experiment, because even the absence of multitooth scintillations would be a challenge to science; indeed, it would imply a manifestation of some new physics, with the proton or its components dropping under the gravitational radius of the daemon, into other dimensions, and so on.

Second, in order to create the BL and, thus, to realize controlled catalysis of light nucleus fusion, one will have to capture a superslow daemon in a carbon fiber. This can be done by different means. Originally one will possibly have to fabricate a piled carbon fiber cloth. The layers of this cloth should alternate with scintillation detector layers. If a scintillation signaling the passage of a slow daemon through the system has appeared, a potential of a few MV should be applied to the cloth with a certain delay to slow down the daemon as it approaches the cloth. On entering a carbon fiber, the slowed-down daemon will start to capture and fuse carbon nuclei. Ejection of electrons will charge the fiber positively to a potential of a few kV, so that if the daemon slides down to the bottom fiber tip it will remain in it in an electrostatically suspended state. Here the daemon will continue to fuse carbon nuclei, with an electronic and a $^{24}$Mg halo surrounding it, i.e., a BL will light up!

Further steps in development of a continuously operating daemon source of power will be a matter of correct, purely engineering solutions. The need of capturing the daemons will possibly no longer arise in the future if we learn how to multiply them (this possibility is again suggested by BL observations, see the end of Sec. II).

Obviously enough, the availability in Man's possession of a compact (a few cm in size), ecologically clean nuclear source of power fuelled by ubiquitous carbon and offering a power output of a few kilowatts would be comparable in its tremendous implications to harnessing the fire, because it would satisfy for centuries the energy demands of Mankind.